\documentclass[12pt]{article}
\usepackage{graphicx} 
\usepackage{epsfig}
\usepackage{amsmath, amsfonts, amsthm,amssymb}
\usepackage{latexsym}
\usepackage{xcolor}
\usepackage{mathrsfs}
\usepackage{natbib}
\usepackage{hyperref}
\usepackage{appendix}
\usepackage{dsfont}
\usepackage[left=3cm,top=4cm,right=3cm,bottom=3cm]{geometry}
\RequirePackage[caption=false]{subfig}
\usepackage{lscape}
\usepackage{dsfont}
\usepackage{rotating}
\usepackage{ulem}
\usepackage{adjustbox}
\usepackage{comment}

\usepackage{mathabx}
\usepackage{calligra}

\graphicspath{{./figs/}}

\newcommand{\de}{\mathrm{d}}
\newcommand{\e}{\mathds{E}}
\newcommand{\R}{\mathds{R}}
\newcommand{\1}{\mathds{1}}
\newcommand{\AM}{\mathbf{A}}
\newcommand{\LM}{\mathbf{L}}
\newcommand{\MM}{\mathbf{M}}
\newcommand{\X}{\Breve{X}}

\DeclareMathOperator{\nach}{ne}

\DeclareMathOperator{\trace}{\mathrm{trace}}

\DeclareMathOperator{\abs}{abs}

\DeclareMathOperator{\tf}{\mathfrak{t}}

\DeclareMathOperator{\diag}{diag}

\DeclareMathOperator{\bo}{\boldsymbol{\Omega}}
\DeclareMathOperator{\bg}{\boldsymbol{\Gamma}}
\graphicspath{{./figs/}}

\begin{document}
\title{\LARGE \bf{Second-order characteristics for spatial point processes with graph-valued marks} }
\maketitle
\begin{center}
{{\bf Matthias Eckardt$^{a}$, Farnaz Ghorbanpour$^{b}$} and {\bf Aila S{\"a}rkk{\"a}$^{c}$}}\\
\noindent $^{\text{a}}$ Chair of Statistics, Humboldt-Universit\"{a}t zu Berlin, Berlin, Germany\\
\noindent $^{\text{b}}$ Department of
Mathematical Sciences, Allameh Tabataba'i
University, Tehran,  Iran\\
\noindent $^{\text{c}}$ Department of Mathematical Sciences, Chalmers University of Technology and University of Gothenburg, Gothenburg, Sweden
\end{center}
\begin{abstract}

The immense progress in data collection and storage capacities have yielded rather complex, challenging spatial event-type data, where each event location is augmented by a non-simple mark. Despite the growing interest in analysing such complex event patterns, the methodology for such analysis is not embedded well in the literature. In particular, the literature lacks statistical methods to analyse marks which are characterised by an inherent relational structure, i.e.\ where the mark is graph-valued. Motivated by epidermal nerve fibre data, we introduce different mark summary characteristics, which investigate the average variation or association between pairs of graph-valued marks, and apply some of the methods to the nerve data.    

\end{abstract}
{\it Keywords:  Epidermal nerve fibres; Frobenius inner product;  Graph metric; Mark variogram, Mark-weighted K-function.}

\maketitle

\section{Introduction}

Marked point processes consist of points in some, typically Euclidean, space (locations) which are equipped with marks in a Polish space. Points can, for instance, represent locations of trees or particles, while marks attached to the points contain some further information on the point, such as species or height of a tree or size of a particle. Marks can be very general, such as sets or vectors, giving, e.g., the shape of a tree crown or a time series of tree diameters.  
Even though the general theory of marked point processes cover very general marks, the literature lacks methods to handle structured marks, such as graphs or shapes and, in practise, whence, their analysis remains a mostly unsolved problem and recent challenge in methodological research.  \cite{Ghorbani2021} give a general framework for functional marked point processes, which covers the case where marks are time series, and introduce some summary statistics, such as the marked $K$ function. \cite{BonneauStoyan2022} model graphs like fracture networks with marked point processes. However, in their set-up, the marks are lengths of line segments, not complete fractures. Here, 
we will include the entire graph (fracture) as a mark. Therefore, the mark itself is characterised by an inherent structure, i.e.\ a graph, and its precise form is completely determined by the edges and nodes of that graph. In what follows, we call any such structured mark graph-valued.
Posing additional complexity on the analysis of the marked pattern, suitable methods are required to not only account for the non-scalar nature but also for the underlying  mathematical structure of the marks. An example of a point pattern with 
graphs as marks we have in mind and will analyse in Section \ref{sec:application} is a nerve pattern, where each nerve tree consists of a base point, branching points, and nerve endings which are interconnected to each other.

Summary statistics, such as mark-correlation function or mark variogram, play an essential role in the preliminary data analysis of marked point patterns when describing possible dependence between points, marks, and points and marks. To estimate such summary statistics from the data where the marks are graphs, one has to be able to compute the distance between two graphs. We suggest here to measure the distance between two graphs by using tools from graph theory. In particular, we introduce different graph mark summary characteristics to investigate the association between graph-valued marks of any pair of points as a function of their interpoint distance. 

The paper is organised as follows. First in Section \ref{Sec:definition}, we give a general definition of graph-valued marks. Then, after reviewing 
commonly used mark summary characteristics for real-valued marks in Section \ref{sec:reals}, we discuss extensions of the mark variogram to graph-valued marks in Section \ref{Sec:gmpp:vario}. Further mark-weighted characteristics are presented in Sections \ref{Sec:gmpp:corr} and \ref{Sec:gmpp:gweight}. Estimation of the proposed summary statistics is discussed in Section  \ref{Sec:gmpp:est} and a simulation study presented in Section \ref{Sec:gmpp:sim}. We apply some of the methods to two epidermal nerve fibre patterns in Section \ref{sec:application}. The final conclusions are given in Section \ref{sec:conclusion}.    

\section{Spatial point processes with graph-valued marks}

\subsection{Definition}
\label{Sec:definition}

Let $X=\lbrace x_i, G(x_i)\rbrace$ denote a marked spatial point process on $\R^d \times \mathds{G}$ with points $x_i$ and associated graph-valued marks $G(x_i)$. At each point $x_i$, $G(x_i)$ is a  tuple  $(V(x_i), E(x_i))$ with vertices $v_s\in V(x_i)$ and edges $e_j \in  E(x_i)\subseteq V(x_i) \times V(x_i)$ and  of order $|V(x_i)|$, i.e. the cardinality of $V(x_i)$.  In general, $G(x_i)$ is assumed to be undirected such that for any $v_s\neq v_t$,  $(v_s,v_t)\in E(x_i)$ whenever $(v_t,v_s)\in E(x_i)$. Further, both the points $x_i$ and the associated marks $G(x_i)$ are assumed to be simple where simplicity of the points means that multiple points are not allowed and simplicity of the marks means that each pair of distinct vertices $v_s$ and $v_t$ is at most once in $E(x_i)$.  The graph $G(x_i)$ is fully determined by its ($|V(x_i)|\times |V(x_i)|)$-dimensional adjacency matrix $\AM(x_i)$ where the elements $A_{st}(x_i)$ are nonzero if the nodes $v_s,v_t\in V(x_i)$ are linked by an edge in $G(x_i)$ and zero otherwise. In this case, $v_s$ and $v_t$ are called adjacent and the set of all adjacent nodes $v_t$ of $v_s$ constitutes  the neighbourhood of $v_s$ denoted by $\nach(v_s)$. Further, for any $v_s\in V(x_i)$, the number of adjacent nodes $v_t$ is quantified by the degree $\deg_s(x_i)$ such that $\deg_s(x_i)=\sum_{t}A_{st}(x_i)$. In general, we restrict to the case where $\deg_s(x_i) >0~\forall~v_s\in V(x_i)$ such that there are no isolated nodes in $G(x_i)$. The degrees can efficiently be stored in the diagonal matrix  $\mathbf{D}(x_i)$ with elements $D_{tt}(x_i)=\deg_t(x_i)$ with associated Laplacian matrix $\LM(x_i)=\mathbf{D}(x_i)-\AM(x_i)$ and its normalized version  $\LM^{\mathrm{norm}}(x_i)=\mathbf{D}^{1/2}(x_i)\LM(x_i)\mathbf{D}^{1/2}(x_i)$ \citep[see e.g.][]{Diestel}.  
We note that under the above definitions, the number of vertices  in the graphs $G(x_i),\ i=1,...,n$,   does not have to be the same for each $i$, i.e.\  $|V(\cdot)|$ may vary over the graph-valued marks $(G(x_1),\ldots, G(x_N))$. 

Further, we assume that $\mathds{G}$ is a complete separable metric space and equipped with $\sigma$-algebra $\mathfrak{G}$.  The $\sigma$-algebra of $X$ will be denoted by $\mathfrak{B}$. Associated with the point process $X$, the observed point pattern and the related ground (i.e.\ unmarked) process will be denoted by ${\bf x}$   and $\X$, respectively. For $\X$, the expected number $\e\left[ N(B)\right]$ of points in  $B$ in Borel set $\mathfrak{B}$ is 
\[
\Breve{\Lambda}(B) = \e\left[ N(B)\right] = \int_B \lambda(u)\de u,
\]
where $\lambda$ is the intensity function of $\X$. Analogously, we write $\Lambda(B\times L)=\e\left[ N(B \times L)\right]$ to denote the expected number of points in $B$ with graph-valued marks in the set $L\in \mathfrak{G}$.  Under stationarity of $X$, i.e.\ when $\lbrace x_i, G(x_i) \rbrace\overset{d}{=}\lbrace x_i+x, G(x_i)\rbrace$ holds for any $x\in\R^2$, $\Lambda(B\times L)$ simplifies to
\[
\Lambda(B \times L)=\lambda\nu(B)\mathcal{G}(L),
\]
where the intensity $\lambda$ is constant, $\mathcal{G}$ denotes the mark distribution,  i.e.\ a probability measure on $\left[\mathds{G}, \mathfrak{G}\right]$, and $\nu(B)$ the area (Lebesgue measure) of $B$. Whence $\lambda\mathcal{G(L)}$ is the intensity of $X$ with respect to $L$. 
Further, the first two moments of the marks are 
\[
\mu_G = \int_\mathds{G} G\mathcal{G}(\de G)
\]
and 
\[
\sigma^2_G =  \int_\mathds{G} (G-\mu_G)^2\mathcal{G}(\de G).
\]
For stationary $X$ and any measurable function $f: \R^2\times \mathds{G} \mapsto \R$, the  expectations can be computed by using the Campbell theorem
\begin{equation}
    \e\left[ \sum^{\neq}_{[x,G]\in X} f(x,G) \right] = \lambda\int_{\mathds{R^2}} \int_{-\infty}^{\infty} f(x,G) \mathcal{G}(G) \de x\de G,
\end{equation}
where the $\sum^{\neq}$ symbol denotes the sum over distinct pairs of points.

\subsection{General framework for summary statistics
}\label{sec:reals}

Before establishing how mark summary characteristics  can be obtained for graph-valued marks, we recall the general framework. We note that although the following restricts to stationary and isotropic processes according to the literature, any of the proposed summary statistics could in principle be generalized to the non-stationary and/ or anisotropic case.  

\subsubsection{Mark-correlation functions
}

For a stationary and isotropic marked spatial point process $\lbrace x_i, m(x_i)\rbrace_{i=1,\ldots, N}$ with points $x_i\in\R^2$ and marks $m(x_i)\in \mathds{M}$ in some Polish space $\mathds{M}$, the average  interrelation between the marks can be analysed by the conditional expectation of some test function $t_f:\mathds{M}\times\mathds{M} \to \R^+$, namely
\begin{eqnarray}\label{eq:tf:corr:ctf:fct}
c_{t_f}(r)
=
\e_{ \circ r}[t_f(m(\circ),m(\mathbf{r}))],
\end{eqnarray}
which itself takes the marks $m(\circ)$ and $m(\mathbf{r})$ at the origin $\circ$ and any alternative point in distance  $||\mathbf{r}||=r>0$ from $\circ$ as arguments \citep{PenttinenandStoyan1989,Schlather2001}. Here, $\e_{\circ r}$ denotes the conditional expectation with respect to the joint distribution $M(\de m(\circ), \de m(\mathbf{r}))$ of the marks $m(\circ)$ and $m(\mathbf{r})$, i.e.\ the two-point mark distribution, given that there are points of the process at the origin $\circ$ and at a point distance $||\mathbf{r}||=r$ away with marks $m(\circ)$ and $m(\mathbf{r})$, respectively. We note that under the independent mark assumption, $M(\de m(\circ), \de m(\mathbf{r}))$ decomposes into the product of the marginals $M(\de m(\circ))M(\de m(\mathbf{r}))$.  Formally, $c_{t_f}(r)$ can be interpreted as the ratio of the functions $\varrho^{(2)}_{\tf_f}(r)$ and $\varrho^{(2)}(r)$, i.e.\ the densities of the second-order factorial moment measures $\alpha_{\tf_f}^{(2)}$ and $\alpha^{(2)}$, respectively, where 
\begin{equation}
\alpha_{\tf_f}^{(2)}(B_1\times B_2 \times L_1 \times L_2)=\e\left[\sum^{\neq}_{\substack{(x_1,m(x_1)),\\(x_2, m(x_2))}\in X}\1_{B_1}(x_1)\1_{B_2}(x_2)\tf_f(m(x_1),m(x_2))\right]     \end{equation}
and
\begin{equation}
\alpha^{(2)}(B_1\times B_2)=\e\left[\sum^{\neq}_{x_1,x_2\in\X}   \1_{B_1}(x_1)\1_{B_2}(x_2)   \right]  
\end{equation}
with $B_1, B_2$ denoting subsets of $\R^2$, $L_1, L_2$ subsets of $\mathds{M}$, and $\1$ is an indicator function. 
Using either the distances or products of the marks  $m(\circ)$ and $m(\mathbf{r})$ as construction principle of the test function $\tf_f$, the computation of  \eqref{eq:tf:corr:ctf:fct} yields different mark summary characteristics with each of these investigating  particular distributional properties of the marks as a function of the interpoint distance $\Vert \mathbf{r}\Vert = r$. The most  common choices for $\tf_f$ from the literature are Stoyan's mark correlation function $k_{mm}(r)$, where 
\begin{equation}\label{eq:markcorr}
\tf_f(m(\circ),m(\mathbf{r}))=m(\circ)\cdot m(\mathbf{r}),
\end{equation}
the $\mathbf{r}$-mark functions $k_{m\bullet}(r)$ resp. $k_{\bullet m}(r)$, 
where 
\begin{equation}\label{eq:rmark}
\tf_f(m(\circ),m(\mathbf{r}))=m(\circ) \text{~resp.~}  \tf_f(m(\circ),m(\mathbf{r}))=m(\mathbf{r}),
\end{equation}
and the mark variogram $\gamma_{mm}(r)$ \citep{cressie93,markvar,Stoyan2000}, where 
\begin{equation}\label{eq:markvario}
\tf_f(m(\circ),m(\mathbf{r}))=0.5\cdot(m(\circ)-m(\mathbf{r}))^2.
\end{equation}
In addition, the so-called mark differentiation function $\Delta_{mm}(r)$ \citep{20113358596, HUI2014125} 
with the test function 
\begin{equation}\label{eq:markdiff}
  \tf_f(m(\circ),m(\mathbf{r}))=1 - \frac{\min \left( m(\circ),m(\mathbf{r}) \right)}{\max \left( m(\circ),m(\mathbf{r}) \right)}  
\end{equation}
is used e.g.\ in forestry. 
Instead of computing  the expectation in \eqref{eq:tf:corr:ctf:fct},  it is sometimes preferable to scale $c_{\tf_f}(r)$ by its expected value when $r\to \infty$, 
\begin{equation}\label{eq:ctf:iid:marks:scalars}
 c_{\tf_f}=\int_{\mathds{M}}\int_{\mathds{M}}\tf_f(m(\circ), m(\mathbf{r}))M(\de m(\circ))M(\de m(\mathbf{r}))   
\end{equation}
leading to the $\tf_f$-correlation function $\kappa_{\tf_f}(r)$,
\begin{eqnarray}\label{eq:tf:corr:fct}
\kappa_{\tf_f}(r)
=
\frac{
c_{\tf_f}(r)
}{
c_{\tf_f}
    }.
\end{eqnarray}
Solving \eqref{eq:ctf:iid:marks:scalars} for the test function \eqref{eq:markcorr} as used in Stoyan's mark correlation function yields the mark mean squared $\mu_m^2$ and, in turn, as $c_{\tf_f}(r)$ is assumed to coincide with $c_{\tf_f}$ under the independent mark assumptions, \eqref{eq:tf:corr:fct}  becomes constantly equal to one. To give an example, Stoyan's mark correlation function $k_{mm}(r)$ and its normalized versions $ \kappa_{mm}(r)$ are obtained by solving 
\[
k_{mm}(r)=\e_{\circ r}\left[m(\circ)\cdot m(\mathbf{r})\right]
\]
and 
\[
\kappa_{mm}(r)=\frac{\e_{\circ r}\left[m(\circ)\cdot m(\mathbf{r})\right]}{\mu_m^2}
\]
where $\Vert\mathbf{r}\Vert=r$ and $\mu_m^2$ follows from solving \eqref{eq:ctf:iid:marks:scalars} for $\tf_f(m(\circ), m(\mathbf{r}))=m(\circ)\cdot m(\mathbf{r})$. 


\subsubsection{Mark-weighted $K$-function
}\label{sec:markweighted}

Whereas the above summary characteristics are constructed explicitly from the test function $\tf_f$, the test functions could also be introduced as a weight into the second-order moment function yielding Penttinen's and Stoyan's mark-weighted $K_{\tf_f}$-function \citep{pettinen1992forest} where  
\begin{equation}\label{eq:weightedK}
 K_{{\tf}_f}(r) = \frac{1}{\lambda c_{{\tf}_{f}}} \e_{\circ r} \left[\sum_{x_i \in \X}\tf_{f} \left( m(\circ),m(x_i) \right) \1\{ \Vert\circ, x_i\Vert \leq r \} \right]
\end{equation}
with $\lambda$ as above and $c_{\tf_f}$ denoting the conditional expectation of the specific test function for $r\to\infty$ of \eqref{eq:ctf:iid:marks:scalars}. This gives the expected number of further points within the distance $r$ from the origin $\circ$ given that there is indeed a point at the origin weighted by the test function under consideration. Under the random labelling assumption, the test function coincides with its expectation and  \eqref{eq:weightedK} reduces to the second-order moment function of the ground process $\X$. For completeness, ignoring the normalising constant  $c_{\tf_f}$  in \eqref{eq:weightedK} yields a mark-weighted reduced second moment function
\[
C_{\tf_f}(r)=\frac{1}{\lambda} \e_{\circ r} \left[\sum_{x_i \in \X}\tf_{f} \left( m(\circ),m(x_i) \right) \1\{ \Vert\circ, x_i\Vert \leq r \} \right]
\]
where the marks are only adjusted for the intensity of the points but not the expected mean of the test function. 

\subsection{Summary statistics for graph-valued marks
}\label{Sec:gmpp:vario}

If the marks are scalar quantities, it is straightforward to compute the Euclidean distance or product between two marks needed to compute the summary statistics in Section \ref{sec:reals}. However, when the marks are graphs Euclidean distance has to be replaced by some other metric. 
We next concern mark summary characteristics for the point process $\lbrace x_i, G(x_i)\rbrace$ on $\R^2\times \mathds{G}$. To this end, let $G(\circ)$ and $G(\mathbf{r})$ denote the graph-valued marks at locations $\circ$ and $\mathbf{r}$ with $\Vert\mathbf{r}\Vert=r$ and denote by $c_{\tf_f}^G(r)$ the conditional expectation of the map $\tf^G_{f}:\mathds{G}\times \mathds{G}\mapsto \R$ given that there are indeed two points at both locations. We note that the conditional expectation is defined with respect to the two-point mark distribution $\mathcal{G}(\de G(\circ)\de G(\mathbf{r}))$, i.e. for any two sets $L_1, L_2$ in $\mathfrak{G}$,  $\mathcal{G}(L_1\times L_2)$ would relate to the probability of observing $G(\circ)\in L_1$ and $G(\mathbf{r})\in L_2$. Different second-order summary characteristics could then be obtained by specifying the test function for $G(\circ)$ and $G(\mathbf{r})$ for distinct pairs of points in distance $\Vert\mathbf{r}\Vert=r$. Using $c_{\tf_f}^G(r)$ and defining $c_{\tf_f}^G=c_{\tf_f}^G(r)(\infty)$ by 
\begin{equation}\label{eq:ctf:iid:marks:graphs}
c_{\tf_f}^G=\int_{\mathds{G}}\int_{\mathds{G}}\tf_f^G(G(\circ), G(\mathbf{r}))\mathcal{G}(\de G)\mathcal{G}(\de G(\mathbf{r}))   
\end{equation}
 the graph $\tf_f^G$-correlation function 
\begin{equation}\label{eq:kappa:graph}
\kappa_{\tf_f}^G(r)=\frac{c_{\tf_f}^G(r)}{c_{\tf_f}^G}
\end{equation}
appears as straightforward extension of \eqref{eq:tf:corr:fct} to the present context. In what follows, particular interest is placed on the derivation of the mark variogram for graph-valued marks. Apart from the mark variogram, we also discuss extensions of Stoyan's mark correlation function,  the $\mathbf{r}$-mark function and the mark differentiation function as well as the mark-weighted $K_{\tf_f}$ function to the present context. 

\subsubsection{Graph mark variogram}

Adapting the test function of \eqref{eq:markvario} to the pair of marks $(G(\circ),G(\mathbf{r}))$, the graph-based analog of the mark variogram, the graph mark variogram $\gamma_G$, can be constructed by taking the conditional expectation of the half-squared graph distance $d_G(\cdot)$ between the marks $G(\circ)$ and $G(\mathbf{r})$,
\begin{equation}
\gamma_G(r)= 0.5\,\e_{\circ, r}\left[(d_G(G(\circ),G(\mathbf{r})))^2\right], r>0.
\end{equation}
However, different from Euclidean distance used in the scalar-valued mark scenario, the distance $d_G(\cdot)$ between the graph-valued entities $G(\circ)$ and $G(\mathbf{r})$ can be  computed in many different ways, with each of them focusing only on certain aspects of similarity. In consequence, the specific choice of the graph metric used in the test function determines the object of interest in the structural analysis and its interpretation \citep[see][for general review]{10.1214/18-AOAS1176,graph:metrics}. In general, $\gamma_G$ could address three different aspects of variation within the marks corresponding to (i) variation in the local graph structure of the nodes or edges, (ii) variation at the mesoscale structure of the graphs over the sets of adjacent nodes, e.g.\ the neighbours, and (iii) the global structure of the entire graphs   (see Table \ref{tab:graphmetrics} in the Supplement for a summary of selected metrics). 

Focusing on the variation in the local graph structure first, suitable local measures are provided by the following three graph metrics.    
For marks $G(\circ)$ and $G(\mathbf{r})$ the (normalized)  Hamming distance $d_H(\cdot)$,
\[
d_H(G(x_i),G(x_j))=\frac{1}{|V(x_i)|(|V(x_i)|-1)}\Vert \AM(x_i)-\AM(x_j)\Vert_1,
\]
and Jaccard distance $d_{J}(\cdot)$,
\[
d_J(G(x_i),G(x_j))=\frac{\Vert\AM(x_i)-\AM(x_j)\Vert_1}{\Vert\AM(x_i)-\AM(x_j)\Vert{_\ddagger}},
\]
with $\Vert\cdot\Vert_1$ and $\Vert\cdot\Vert_{\ddagger}$ denoting the $L_1$ and the nuclear norm, i.e.\ sum of the singular values, of a matrix, respectively, are particular instances of the graph edit distance which quantify the number of necessary edge addition or deletion operations to transform $G(\circ)$ into  $G(\mathbf{r})$ \citep[see][for detailed review on different graph edit  distances]{Gao2010}. Requiring correspondence of the nodes over all marks $(G(x_1),\ldots, G(x_n))$, both of these metrics investigate the pairwise variation of the adjacency matrices $\AM(\circ)$ and $\AM(\mathbf{r})$ and are useful tools to decide on the pairwise local variation of the nodes among the marks. Similarly, the  Frobenius norm distance $d_F(\cdot)$,
\[
d_F(G(x_i),G(x_j))=\sqrt{\trace((\AM(x_i)-\AM(x_j))^{\top}(\AM(x_i)-\AM(x_j)))},
\]
can be used to investigate the similarity of the relational systems among the nodes between the pairs $G(\circ)$ and $G(\mathbf{r})$.  
In contrast, variation at the mesoscale structure of the marks can be investigated using the 
matrix distance $d_\MM(\cdot)$,
\[
d_\MM(G(x_i),G(x_j))=\mathrm{abs}(\MM(x_i) - \MM(x_j)) ,
\]
and the Matusita difference  $d_\mathbf{W}(\cdot)$,
\[
d_\mathbf{W}(G(x_i),G(x_j))=\sqrt{\left(\sum_{i,j}\sqrt{\mathbf{W}(x_i)}-\sqrt{\mathbf{W}(x_j)}\right)^2},
\]
where 
$\MM(x_i)$ is defined through some  distance $\delta: V\times V \mapsto \R$ with elements $M_{st}=\delta(v_s,v_t) \in V(x_i)$ of $G(x_i)$ and $\mathbf{W}(x_i)$ is the fast belief propagation matrix \citep{10.1145/2824443} defined by 
\[
\mathbf{W}(x_i)=\left[\mathbf{I}+\epsilon^2\mathbf{D}-\epsilon\AM(x_i) \right]^{-1}
\]
which can also be formulated as the power series $\mathbf{W}(x_i)=\mathbf{I}+\epsilon\AM(x_i) + \sum_{k} \epsilon^k(\AM^k(x_i)-\mathbf{D}(x_i))$. Here, $\epsilon^k$, $k=2,3,4,...$ is a weight for neighbours separated by a path of length $k$, $\mathbf{I}$ the identity matrix and $\mathbf{D}$ the degree matrix. For $\MM_{st}$, common choices of $\delta$ include the shortest-path distance \citep{1570854175170619520} which considers only single paths between the vertices, the effective graph resistance \citep{ELLENS20112491} which considers all possible paths between the vertices, the random walk distance \citep{10.1145/511446.511513}, and the resistance-perturbation distance \citep{MONNIG2018347} which measures the changes in connectivity between two graphs based on the effective graph resistance, which can be written in terms of Laplacian eigenvalues.  

Several metrics which could be used to capture differences in the global structure of the marks $G(\circ)$ and $G(\mathbf{r})$ have also been proposed in the literature. A suitable first class of metrics are the so-called spectral distances including the adjacency spectral distance  $d_\AM(\cdot)$, 
\[
 d_\AM(G(x_i),G(x_j))=\sqrt{\left(\sum_{l=1}^n\omega^{\AM(x_i)}_l-\omega^{\AM(x_j)}_l\right)^2}
\]
the Laplacian spectral distance $d_\LM(\cdot)$,
\[
d_\LM(G(x_i),G(x_j))=\sqrt{\left(\sum_{l=1}^n\omega^{\LM(x_i)}_l-\omega^{\LM(x_j)}_l\right)^2}
\]
and the normalized Laplacian spectral distance  $d_\LM^{\mathrm{norm}}(\cdot)$ \citep{10.3389/fncom.2013.00189}
\[
d_\LM^{\mathrm{norm}}G(x_i),G(x_j))=\sqrt{\left(\sum_{l=1}^n\omega^{\LM^{\mathrm{norm}}(x_i)}_l-\omega^{\LM^{\mathrm{norm}}(x_j)}_l\right)^2}.  
\]
The spectral distances are based on the eigendecomposition  of e.g.\ the adjacency $\AM(x_i)$ or Laplacian $\LM(x_i)$ matrices of $G(x_i)$ into $\bg(x_i)\bo(x_i)\bg^{\top}(x_i)$ where $\bo(x_i)=\diag(\omega_1(x_i),\ldots,\omega_{|V|}(x_i))$ with ordered eigenvalues $\omega_1(x_i)\geq \omega_2(x_i) \geq\ldots\geq \omega_{|V|}(x_i)$ and $\bg(x_i)=(\gamma_1(x_i) | \gamma_2(x_i) |\ldots|\gamma_{|V|}(x_i))$ is a matrix with ordered eigenvectors \citep[see][for a detailed treatment]{WILSON20082833,chungspectral}. We note that the above ordering of the eigenvalues only applies to the eigendecompostion of the adjacency matrix while the eigendecomposition of the (normalized) Laplacian yields a reversed ordering with $\omega_1(x_i)\leq \omega_2(x_i) \leq\ldots\leq \omega_{|V|}(x_i)$. A different metric, 
the eigenspectrum distribution distance $d_{\varrho}(\cdot)$ \citep{GU201630} derives from a continuous spectral distribution which (using a Gaussian kernel) takes the following form 
\[
\varrho(x_i)(a) = \frac{1}{|V(x_i)|}\sum^{|V(x_i)|-1}_{l=0}\frac{1}{\sqrt{2\pi \sigma^2}}\exp\left\{-\frac{a-\omega_l(x_i)}{2\sigma^2}\right\}
\]
and is defined by 
\[
d_{\varrho}(G(x_i),G(x_j))=\int |\varrho(x_i)(a) -\varrho(x_j)(a)| \de a.
\]
Similarly to the graph edit distance, the spanning tree distance $d_{ST}(\cdot)$ \citep{10.1214/18-AOAS1176}, 
\[
d_{ST}(G(x_i),G(x_j))=|\log(\mathcal{T}(x_i))-\log(\mathcal{T}(x_j))|,
\]
quantifies the number of spanning tree  construction or deconstruction operations needed to transform the first into the second graph where for each mark $G(x_i)$ the number of spanning trees $\mathcal{T}(x_i)$  derives from the eigenvalues $\omega_l(x_i)$ of the Laplacian $\LM(x_i)$ through $\mathcal{T}(x_i)=|V(x_i)|^{-1}\prod^{|V(x_i)|-1}_{l=1} \omega_l(x_i)$.  Two alternative useful global graph metrics are the the Ipsen-Mikhailov distance $d_{IM}(\cdot)$,
\begin{equation}
d_{IM}(G(x_i),G(x_j))=\sqrt{\left(\int_0^{\infty}\varrho(\theta,\xi)(G(x_i))-\varrho(\theta,\xi)(G(x_j))\right)^2\de \theta}\label{formula:ipsenmikhaolov}
\end{equation}
and the Hamming-Ipsen-Mikhailov distance $d^{\xi}_{HIM}(\cdot)$ \citep{HIM},
\[
d^{\xi}_{HIM}(G(x_i),G(x_j))=\frac{1}{\sqrt{1+\xi}}\sqrt{\mathbf{I}\MM^2+H^2}.
\]
Also considering the distances between the spectral densities of the Laplacians of two graphs, the Ipsen-Mikhaolov distance $d_{IM}$ is based on the squares of the vibrational frequencies $\theta_l$ where $\theta_0=\omega_0=0$ and $\theta_l^2=\omega_l$ and is defined by 
\[
d_{IM}(G(x_i),G(x_j))=\sqrt{\left(\int_0^{\infty}\varrho(\theta,\xi)(G(x_i))-\varrho(\theta,\xi)(G(x_j))\right)^2\de \theta}
\]
where $\xi$ is a parameter common to all $\theta$ and needs to be determined  based on the data.

We note that all of the above distances can in general also be applied to graph-valued marks $(G(x_1),\ldots,G(x_N))$ with varying order, i.e.\ when $|V(x_i)|\neq |V(x_j)|$, by introducing artificial nodes $v_p$ into the adjacency matrices where $A_{pt}(x_i)=0$ for all $v_t\in V(x_i)$ to obtain equally sized adjacency matrices of dimension $\max(|V(x_i)|)\times \max(|V(x_i)|)$. Alternatively, the normalized Laplacian distance and distances for metric spaces such as the Gromov-Hausdorff distance \citep{berger2003}, the Wasserstein distance \citep{villani2003topics}, the Hellinger distance \citep{Hellinger1909}, or the Fr{\'e}chet distance \citep{frechetdist:graph} can be applied.     

\subsubsection{Other graph mark correlation functions 
}\label{Sec:gmpp:corr}

The mark correlation function, $\mathbf{r}$-mark function and the mark differentiation function of Section \ref{sec:reals} can also be extended to graph-valued marks by computing the graph norm $\Vert\cdot\Vert_G$ or graph inner product $\langle\cdot,\cdot\rangle_G$ of either the adjacency matrix $\AM(x_i)$, Laplacian matrix $\LM(x_i)$, or the matrices $\mathbf{M}(x_i)$ and $\mathbf{W}(x_i)$. Denoting by $\nabla_G(x_i)$ any of the above ($|V(x_i)|\times|V(x_i)|$)-dimensional matrix representation of the graph-valued mark $G(x_i)$, the graph mark correlation $k_{GG}$ is defined as 
\begin{equation}\label{eq:graphcorStoyan}
k_{GG}(r)=\e_{\circ, \mathbf{r}}\left[\langle\nabla_G(\circ),\nabla_G(\mathbf{r})\rangle_G\right]\end{equation}
and quantifies the average product of any two matrices $\nabla_G$ as a function of the distance $r$. Suitable graph inner products  include e.g.\ the Frobenius  inner product $\langle \cdot, \cdot \rangle_F$ which directly extends the scalar product of \eqref{eq:markcorr} to matrices. For $\nabla_G(x_i)=\AM(x_i)$ and a binary specification of the adjacency matrix,  
 $\langle \AM(x_i), \AM(x_j)\rangle_F$ corresponds to the number of matching edges between the graphs $G(x_i)$ and $G(x_j)$, i.e. if the vertices $(v_s, v_t)$ are in both $E(x_i)$ and $E(x_j)$.
Likewise, applying the graph norm $\Vert\cdot\Vert_G$ yields the graph $\mathbf{r}$-mark functions $k_{G\bullet}(r)$ and $k_{\bullet G}(r)$, where 
\begin{equation}\label{eq:graphRcor}
k_{G\bullet}(r)=\e_{\circ, \mathbf{r}}\left[\Vert\nabla_G(\circ)\Vert_G\right]\text{~and~} k_{\bullet G}(r)=\e_{\circ, \mathbf{r}}\left[\Vert\nabla_G(\mathbf{r})\Vert_G\right].  \end{equation}
Again, specifying $\Vert\cdot\Vert_G$ as the Froebenius norm $\Vert\cdot\Vert_F$ of the matrix $\nabla_G(\circ)$ or $\nabla_G(\mathbf{r})$ where  
\begin{equation}\label{eq:graphRcorr}
\Vert \nabla_G(\circ)\Vert_F=\sqrt{\sum_{s=1}^{|V(\circ)|}\sum_{t=1}^{|V(\circ)|}\nabla_G(st)(\circ)}
\end{equation}
the graph $\mathbf{r}$-mark function becomes 
\[
k_{G\bullet}(r)=\e_{\circ, \mathbf{r}}\left[\Vert\nabla_G(\circ)\Vert_F\right]\text{~and~} k_{\bullet G}(r)=\e_{\circ, \mathbf{r}}\left[\Vert\nabla_G(\mathbf{r})\Vert_F\right].  
\]
Assuming a binary specification of $\AM(x_i)$ and setting $\nabla_G(\circ)=\AM(\circ)$,  $k_{G\bullet}(r)$ quantifies the average number of edges of the first point at distance $r$ for any pair of distinct points conditional on the presence of points at both locations. Further suitable graph norms include the entrywise $p$-norm $\Vert \nabla_G \Vert_{(p)}=(\sum_{s,t}|\nabla_G({s,t})|)^{1/p}$, the $L_p$-operator norm $\Vert \nabla_G \Vert_p$ and the absolute $L_p$-operator norm $\Vert \nabla_G \Vert_{|p|}=\Vert \abs(\nabla_G) \Vert_p$ where $\abs$ refers to the entrywise absolute
values of its argument \citep[see][for a general review of different graph norms]{gervens_et_al:LIPIcs.MFCS.2022.52}. 
Finally, extending the mark differentiation function of \eqref{eq:markdiff} to the present context, the graph mark differentiation function $\Delta_{GG}(r)$ is defined by 
\begin{equation}\label{eq:graphdiff}
\Delta_{GG}(r)=\e_{\circ\mathbf{r}}\left[1-\frac{\sum_{st}\min(\nabla_G(st)(\circ), \nabla_G(st)(\mathbf{r}))}{\sum_{st}\max(\nabla_G(st)(\circ), \nabla_G(st)(\mathbf{r}))}\right].    
\end{equation}

\subsubsection{Graph-weighted $K$-function
}\label{Sec:gmpp:gweight}

Apart from the second-order mark characteristics, all the above test functions can be used to extend the mark-weighted $K_{\tf_f}$ function of Section \ref{sec:markweighted} to a graph-weighted version. Substituting $\tf^G_{f}$ for $\tf_f$ in \eqref{eq:weightedK} leads to the graph-weighted $K^G_{\tf_f}$ function,
\begin{equation}\label{eq:graphKtf}
 K^G_{{\tf}_f}(r) = \frac{1}{\lambda c^G_{{\tf}_{f}}} \e_{\circ r} \left[\sum_{x_j \in \X}\tf_{f} \left( G(\circ),G(x_j) \right) \1\{ \Vert\circ, x_j\Vert \leq r \}|x_j\in X \right],
\end{equation}
which translates into the graph-weighted $L^G_{\tf_f}$ analogous to Section \ref{sec:markweighted}. Using the graph matrices $\nabla_G(\circ)$ and $\nabla_G(x_j)$ as arguments and choosing the Frobenius inner product $\langle \cdot,\cdot \rangle_F$ as test function corresponding to the graph mark correlation, \eqref{eq:graphKtf} becomes the $K_{GG}$ function,
\begin{equation}\label{eq:graphKGG}
 K_{GG}(r) = \frac{1}{\lambda c^G_{{\tf}_{f}}} \e_{\circ r}\left[\sum_{x_j \in \X}\langle \nabla_G(\circ), \nabla_G(x_j)\rangle_F \1\{ \Vert\circ, x_j\Vert \leq r \}|x_j\in X\right].
\end{equation}
where $c^G_{{\tf}_{f}}$ follows from the solution of 
\[
c^G_{{\tf}_{f}}=\int_{\mathds{G}}\int_{\mathds{G}}\langle \nabla_G(\circ), \nabla_G(x_i)\rangle_F \mathcal{G}(\de G)\mathcal{G}(\de G(\mathbf{r})).  
\]
As in the scalar case, \eqref{eq:graphKGG} can be interpreted as the expected number of further points within distance $r$ of $\circ$  weighted by the product of the corresponding graph matrices at these two locations. The exact interpretation of the characteristics depends on the specification of the graph matrix $\nabla_G(x_i)$ and test functions under consideration.       

\subsection{Estimation}\label{Sec:gmpp:est}

Next we outline the estimation of the graph summary characteristics from the point pattern $\mathbf{x}$ collected within an observation window $W\subset \R^2$. Using the representation of $c_{\tf_f}^G(r)$ as the ratio of two product density functions $\varrho^{G,(2)}_{\tf_f}(r)$ and $\varrho^{(2)}(r)$, all the above second order mark summery characteristics can be achieved by computing the ratio of the estimated densities $\widehat{\varrho^{G,(2)}_{\tf_f}}(r)$ and $\widehat{\varrho^{(2)}_{\tf_f}}(r)$ where 
\begin{equation}\label{eq:varrhoG}
\widehat{\varrho^{G,(2)}_{\tf_f}}(r)=\frac{1}{2\pi r \nu(W)}\sum^{\neq}_{x_i, x_j\in W}
\tf^G_f(G(x_i), G(x_j))\mathfrak{K}_b(\Vert x_i-x_j\Vert-r)e(x_i,\Vert x_i-x_j),  
\end{equation}
and
\begin{equation}\label{eq:varrho:classic} 
\widehat{\varrho^{(2)}}(r)=\frac{1}{2\pi r \nu(W)}\sum^{\neq}_{x_i, x_j\in W}
\mathfrak{K}_b(\Vert x_i-x_j\Vert-r)e(x_i,\Vert x_i-x_j) 
\end{equation}
with $\mathfrak{K}_b(\cdot)$ denoting a kernel function of bandwidth $b$, $e(\cdot)$ an edge correction factor and $\nu(W)$ the area of the observation window $W$. The desired characteristics is then obtained through the specific choice of $\tf_f^G$ in \eqref{eq:varrhoG}.  Taking e.g.\ the Frobenius inner product $\langle G(x_i), G(x_j)\rangle_F$ as $\tf_f^G$ and computing $\widehat{c_{\tf_f}^G}(r)$  by dividing \eqref{eq:varrhoG}  by \eqref{eq:varrho:classic} yields the graph mark correlation function $k_{GG}(r)$. By applying the Campbell-Theorem  to \eqref{eq:varrho:classic} it can be shown that $\widehat{\varrho}^{(2)}(r)$ is an unbiased estimator of $\varrho^{(2)}(r)$ as the bandwidth $b\rightarrow 0$   \citep{Illian2008, Chiu2013}. Similarly, extending the Campbell-Theorem to the marked case \citep{Daley2003} $\e\left[\widehat{\varrho^{G,(2)}_{\tf_f}}(r)\right]\rightarrow \varrho^{G,(2)}_{\tf_f}(r)$ when $b$ tends to zero. In consequence, the ratio of both terms yields a ratio-unbiased estimator. In particular, as for both $\varrho^{G,(2)}_{\tf_f}(r)$ and $\varrho^{(2)}(r)$ the same estimation principle is applied,   potential edge effects can be ignored in the formulation of the above second order mark summery characteristics  \citep[see][for detailed discussion]{Illian2008}.

Further, normalized graph summary characteristics can be calculated by estimating the $\tf_f^G$-correlation function of \eqref{eq:kappa:graph}  by 
\[
\widehat{\kappa_{\tf_f}}(r)=\left.\frac{\widehat{\varrho^{G,(2)}_{\tf_f}}(r)}{\widehat{\varrho^{(2)}}(r)}\right/\widehat{c^G_{\tf_f}}
\]
where the normalising factor $c_{\tf_f}^G$ can be estimated from the graph-valued marks $G(x_1),\ldots, G(x_n)$ as 
\[
\widehat{c^G_{\tf_f}}=\frac{1}{|V|^2}\sum_{i=1}^n\sum^n_{j=1}\tf_f(G(x_i), G(x_j))
\]
with $|V|=|V(x_i)|$ for all $x_i$ in $\mathbf{x}$. 

\section{Simulation study}\label{Sec:gmpp:sim}

To evaluate the performance of the proposed summary characteristics in the presence and absence of spatially dependent marks, we first simulated  a completely spatially random point pattern with intensity $\lambda=$ $n= 110$ points using the \texttt{spatstat} package \citep{spatstat} in R \citep{Rcore}. In addition, we also simulated a Strauss point process \citep{rStrauss} with intensity parameter $ \beta=160$, interaction parameter $\gamma=0.3$ and interaction radius $R = 0.05$ yielding $n= 98$ points and a modified Thomas process \citep{rThomas} with intensity $\lambda_p=8$ for the parents points, scale parameter $\sigma= 0.45$ and mean number of points per cluster $\mu=11$ yielding $n=101$ points to investigate potential effects of inhibition or clustering of the points on graph mark summary characteristics.  All patterns were simulated in the unit square. Next, to construct a graph-valued mark for each point location we considered a random graph specification $G(n_v,p)=(V,E)$ according to the Erd{\"o}s-R{\'e}nyi model with $n_V$  vertices provided by the \texttt{igraph} package \citep{igraph}. In this model, each pair of nodes is connected by an edge  with probability $\mathds{P}(\lbrace v_s,v_t \rbrace\in E(x_i)\vert v_s\neq v_t)=p,\ 0<p<1$, independently of all the other pairs. 
To investigate the effect of the number of vertices, we consider two different graph scenarios setting $n_V= 25$ and $n_V= 5$. For the independent mark scenario we considered an Erd{\"o}s-R{\'e}nyi model with  $p=0.5$. Next, to impose some spatial structure among the marks, we computed for each point $x_i$ the distance to the boundary $\mathfrak{b}(x_i)$ of the observation window and specified the corresponding random graph setting $p=\mathfrak{b}(\mathbf{x_i})$.  Note that $\mathfrak{b}(x_i)$ is always less than one since the simulation window is the unit square. 

\begin{figure}
    \centering
    \includegraphics[width=0.85\textwidth]
    {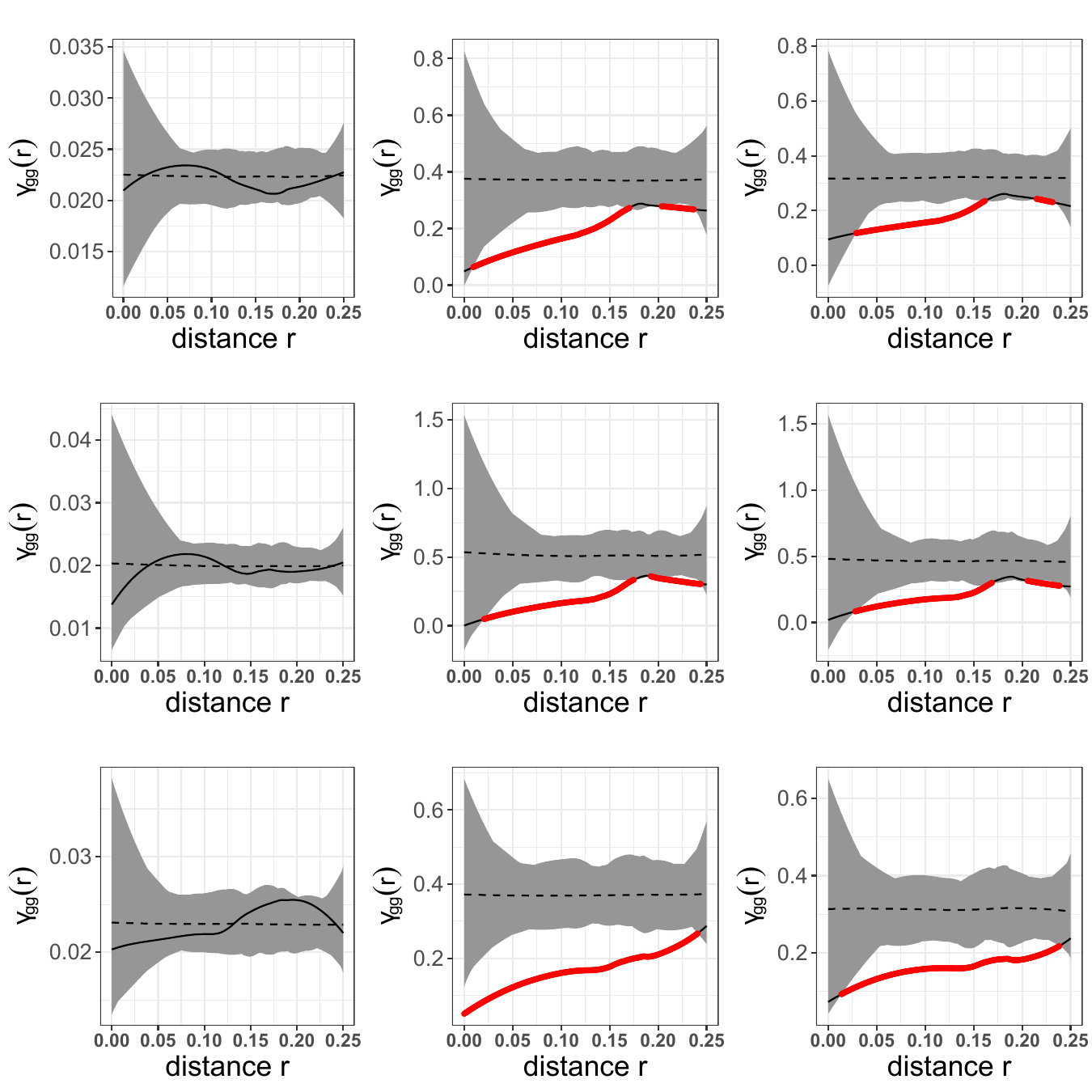}
    \caption{Graph mark variograms for simulated data from a homogenous Poisson process (top row), Strauss process (central row), and modified Thomas process (bottom row) with Erd{\"o}s-R{\'e}nyi graphs $G(n_V,p)$ with $n_V = 25$ as marks and 95\% global envelopes based on 500 random permutations of marks. From left to right: (i) graph mark variogram for $G(n_V,p)$ with  $p=0.5$, (ii) graph mark variogram for $G(n_V,p)$ with $p$ determined by the distance of the points to the boundary of the observation window with binary specified adjacency matrices, and (iii) graph mark variogram for $G(n_V,p)$ with $p$ determined by the distance of the points to the boundary of the observation window and nonzero entities of the adjacency matrices determined by the node degrees.}
    \label{fig:SimRes}
\end{figure}

In addition to the binary specified adjacency matrices under the above setting, we additionally considered the case where $A_{s,t}(x_i)=D_{s,t}(x_i)$, where $D_{s,t}$'s are the elements of the diagonal degree matrix.  We used the Ipsen-Mikhaolov distance \eqref{formula:ipsenmikhaolov} to capture differences in the global structure of the marks. Using all three data scenarios as input, we computed the graph mark variogram and 95\% global envelopes based on the extreme rank length (ER) measure \citep{mari1, MrkvickaEtal2020} and 500 simulations under the random labeling hypothesis using the \texttt{GET} package \citep{GETpack}. The results and the simulated point patterns are shown in Figure \ref{fig:SimRes}. 
For the independent mark scenario depicted in the left panel of Figure $\ref{fig:SimRes}$, the estimated graph mark variogram is completely covered by the global 95\% envelopes corresponding to the absence of any mark correlation in the data. Both alternative plots highligh a clear deviation from the global 95\% envelopes  showing that the marks close together are more similar than the marks further apart. The underlying spatial structure does not affect the results much. 

The results for the second graph scenario with $n_V=5$, which corresponds to the graph structure in the ENF data, are depicted in Figure \ref{fig:SimResNv5} of the Supplement.   They are very similar to the results with larger graphs.

\section{Application to epidermal nerve fibre data}\label{sec:application}

\begin{figure}
    \centering
    \includegraphics[width=0.70\textwidth]
    {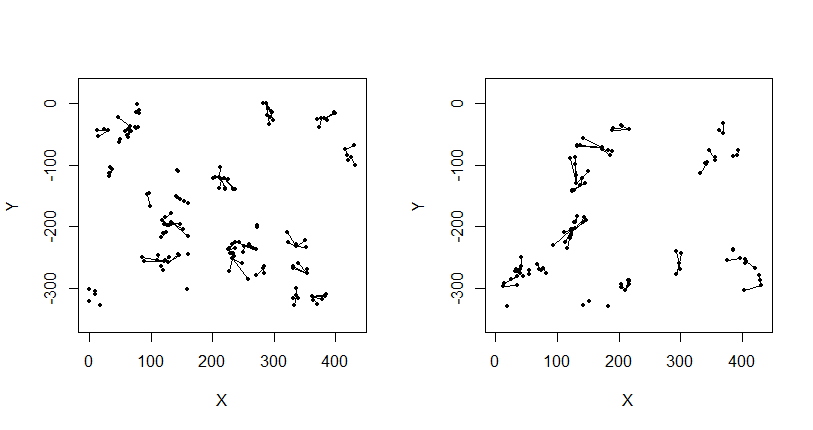}\hskip1cm
\includegraphics[scale=.75]{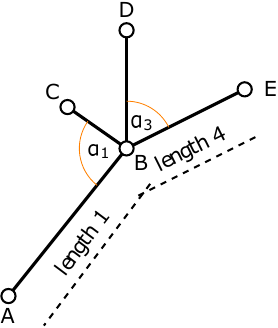}
    \caption{An ENF pattern from a healthy control (left), a pattern from a subject with mild diabetic neuropathy (middle), and a schematic plot of an individual nerve tree (right). Only the base and end points are included in the ENF patterns.   }
    \label{fig:ENFPattern}
\end{figure}

As an example of data consisting of point locations and graph marks, we consider epidermal nerve fiber (ENF) patterns taken from 
two patients using suction skin blister biopsy. The samples were first immunostained and then traced using confocal microscopy, see  \citep{Kennedy1993, Kennedy1999, Wendelschafer2005}, and \citep{Panoutsopoulou2009} for more details of the staining and imaging. The data are in the form of coordinates of the locations of (i) points, where the nerves enter the epidermis, called base points, (ii) points, where the nerve fibers branch, called branching points, and (iii) termination points of the individual nerve fibers, called end points. Therefore, each nerve is a tree like structure consisting of base, branching, and end points which are interconnected. When the ENF coverage across the skin is of interest, it is reasonable to concentrate on 2D projections (330 $\mu$m × 432 $\mu$m) of the 3D patterns.
The ENF patterns are regarded as realizations of marked point processes, where the locations of the base points are points and the entire nerve trees consisting of the branching and end points marks. Two ENF patterns are shown in Figure \ref{fig:ENFPattern} (left, middle) together with the structure of a nerve tree (right). 

It has been reported that patients suffering from peripheral neuropathy, such as diabetic neuropathy, have reduced coverage of the epidermis by ENFs due to a smaller number of ENFs per surface area and reduced summed length of all ENFs per volume compared to healthy controls \cite{Kennedy1996}. Furthermore, the resulting nerve patterns tend to be more clustered than healthy patterns \cite{Kennedy1999}. It was demonstrated in \cite{Konstantinou2023} that removing the most isolated nerve trees in healthy patterns results in patterns similar to neuropathy patterns. However, not only the locations of the nerves but also the tree (mark) structure of them can change due to neuropathy. Therefore, possible changes in the mark structure might give important additional information on how the ENF structure is affected by neuropathy.

%
Here, we concentrate on two patterns, one from a healthy control and one from a subject with mild diabetic neuropathy, shown in Figure \ref{fig:ENFPattern} to illustrate the possible advantage of using of our graph mark summary statistics. For simplicity, we restrict ourselves to one of the mark summary statistics, the mark variogram. First, in Section \ref{sec:RealMarks}, we estimate the mark variogram based on some real valued marks 
and then in Section \ref{sec:graphmarks}, we estimate the graph mark variogram by using the whole tree structure (graph) as the mark. 

\subsection{Mark variogram with real-valued marks}
\label{sec:RealMarks}

\begin{figure}
    \centering
    \includegraphics[width=0.950\textwidth]
    {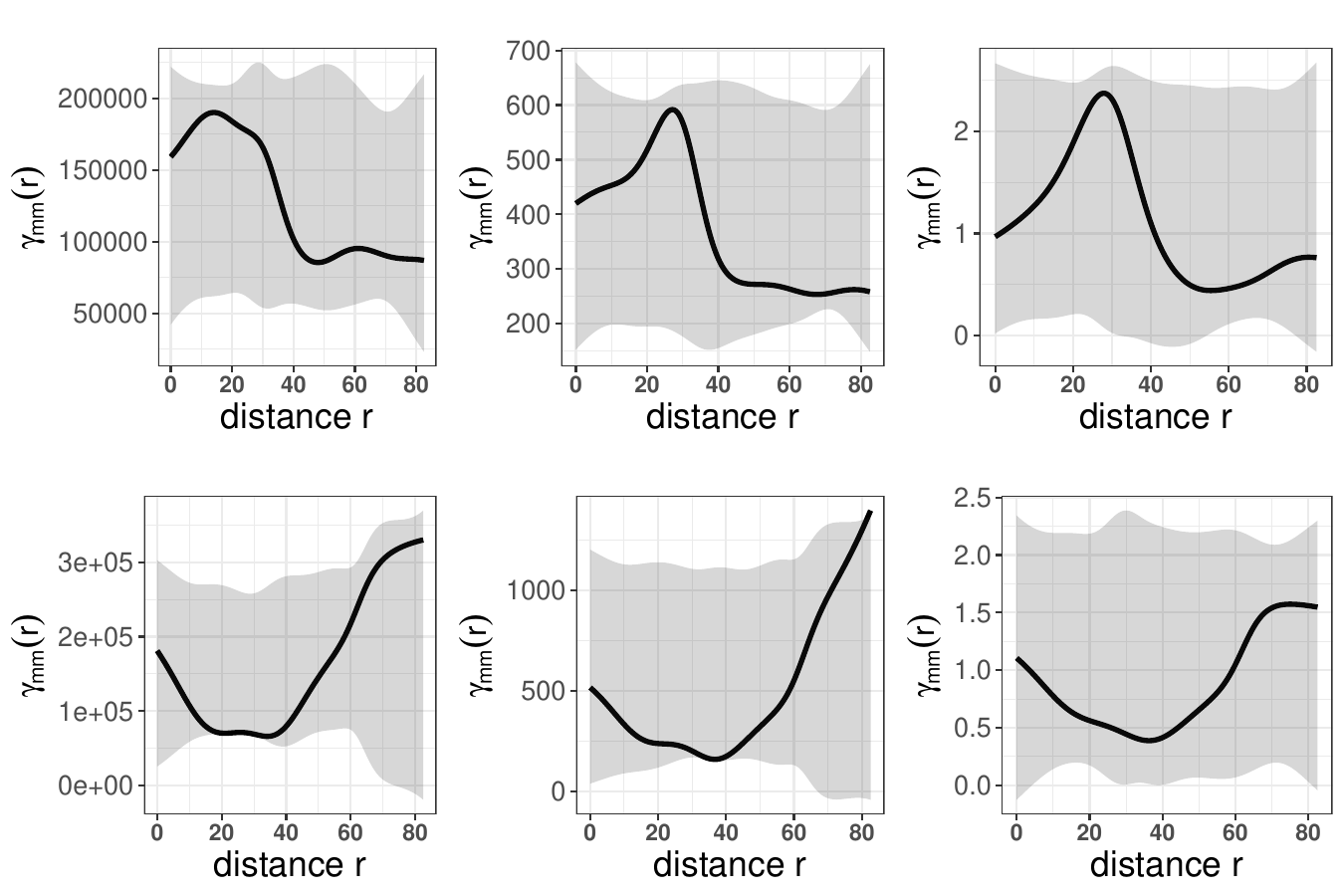}
    \caption{
    The mark variogram based on the area of the reactive territory (left), length of the tree (middle), and order of the tree (right) together with global envelopes based on random labelling for the healthy (top) and neuropathy (bottom) ENF samples shown in Figure \ref{fig:ENFPattern}.}
    \label{fig:SummariesRealMarks}
\end{figure}

For the ENF data, we consider three real-valued marks which have earlier been considered in the literature: (i) the area of the reactive territory, i.e.\ the smallest disc that covers all the base, branching, and end points \citep{Ghorbanpour2021}, (ii) the length of the tree, i.e.\ sum of all branch lengths \citep{Andersson2018}, and (iii) the order of the tree, i.e.\ number of branching layers \citep{Andersson2018}.
We study the ENF patterns given in Figure \ref{fig:ENFPattern} and estimate the mark variogram  based on these three different marks. The results can be seen in Figure \ref{fig:SummariesRealMarks}. The mark variograms based on the area of the reactive territory, length of the tree, and order of the nerve tree do not indicate any mark correlation in these two ENF samples.

\subsection{Mark variogram with graph-valued marks}
\label{sec:graphmarks}

Instead of scalars derived from the graph, we now concern the relational structure of the entire graph as object of our analysis.  
To account for the additional graph-based information in our analysis, we consider graph-based characteristics (weights)  instead of a binary specification of the adjacency matrix $\AM(x_i)$ yielding a weighted adjacency matrix $\AM^{\eta}(x_i)$ with elements 
\begin{equation}
    A_{st}^{\eta}(x_i) = \begin{cases}
        \eta_{st}(x_i), & \mathrm{if~} (v_s,v_t)\in E(x_i)\\
        0, & \mathrm{otherwise}.
    \end{cases}
\end{equation}
In particular, we consider three different specifications of the weight, namely the segment length $\eta_{st}^{(1)}(x_i)$ between vertices $s$ and $t$ in $G(x_i)$, the segment angle of $G(x_i)$ 
$\eta_{st}^{(2)}(x_i)$, and $\eta_{st}^{(3)}(x_i)$, which was constructed as the sum of the corresponding segment lengths and angles (i.e. $\eta_{st}^{(1)}(x_i)$ and $\eta_{st}^{(1)}(x_i)$). Since the mark variogram with real-valued marks did not show any significant mark correlation, we are particularly interested in whether there is  variation in the global structure of the entire graphs. To this end, we chose the Ipsen-Mikhailov distance as the graph metric. We note that $\AM^{\eta}(x_i)$ also affects the off-diagonal elements in $\LM(x_i)$ as $\LM(x_i)=\mathbf{D}(x_i)-\AM^{\eta}(x_i)$ and, in turn, $\LM^{\mathrm{norm}}(x_i)$. 

To estimate the graph based weights, the nodes of the nerve trees are ordered starting from the base (node A in Figure \ref{fig:ENFPattern}), then the first branching point (B), and the nodes connected to the first branching point (C, D, E) from left to right. The next branching layer, if it exists, is ordered similarly. Ordering of the branch lengths follows the same principle. The angles are defined as the angles between the branches connected to each branching point, again from left to right. 

\begin{figure}
\centering
    \includegraphics[width=0.950\textwidth]
    {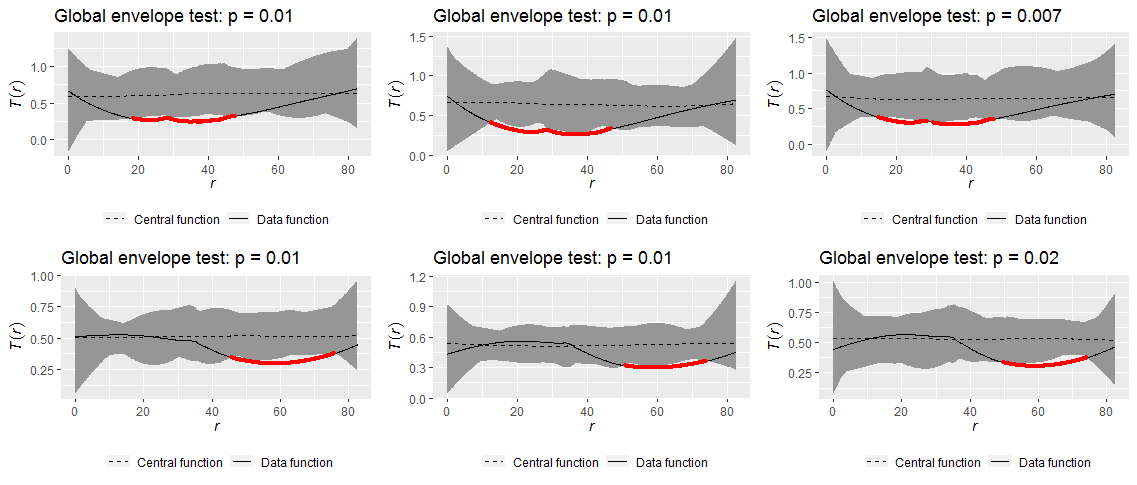}

    
    \caption{The graph based mark variogram 
    based on the segment length (left), segment angle (middle), and the combination of the two (right) together with global envelopes based on random labelling for the healthy (top) and neuropathy (bottom) ENF samples shown in Figure \ref{fig:ENFPattern}.}
    \label{fig:SummariesGraphMarks}
\end{figure}

The estimated mark variograms based on the three weights $\eta_{st}^{(1)}(x_i)$, $\eta_{st}^{(2)}(x_i)$, and $\eta_{st}^{(3)}(x_i)$, for the two ENF samples are shown in 
Figure \ref{fig:SummariesGraphMarks}. All three weights, the one based on the segment lengths, the one based on the angles, and the combination of the two, give similar results.  The average variation of the nerve trees for any pair of points at distances around 20-40 $\mu$m in the healthy sample (top row), and around 50-70 $\mu$m in the neuropathy sample (bottom row) are significantly lower as expected under the independent mark assumption.  At these distances, the ENF tree structures for any pair of points are more similar than expected under random labeling.

\section{Discussion}\label{sec:conclusion}

This paper establishes various mark summary characteristics to investigate the behaviour of the marks within highly complex mark settings, namely when the marks are graphs. 
The summary characteristics are constructed through generalised versions of the classical test functions, such as mark variogram, using different types of graph metrics. By choosing an appropriate distance metric, 
the structure among the marks can be investigated at a local, meso, or global scale. Further, since
the characteristics are generalisations of well-known test functions, they have 
analogue interpretations as the corresponding classical characteristics. To illustrate the new methodology, we applied it to study mark variation in two epidermal nerve fiber patterns, one from a healthy control and one from a subject with mild diabetic neuropathy, where the marks are tree (graph) structures. For these particular patterns, the graph based mark variogram was able to reveal some mark correlation 
between the nerve trees which was not detected by simple real-valued marks indicating that the new characteristics, indeed, can be useful. Furthermore, our results show a difference in the scale of mark correlation between the healthy and neuropathy patterns. However, based on this simple analysis, the result cannot be generalized to concern larger groups of healthy and neuropathy subjects and a much larger study would be needed to make any further conclusions.

\newpage
\section*{Acknowledgments}

The authors gratefully acknowledge financial support through the German Research Association. Matthias Eckardt was funded by the Walter Benjamin grant 467634837 from the German Research Foundation. In addition, Aila S\"arkk\"a would like to thank for the financial support from the Royal Society of Arts and Sciences in Gothenburg.

  \bibliographystyle{biom} 
 \bibliography{gmpp}


\section*{Supporting Information}
\appendix


\section*{Supplementary material}

\subsection{Selected graph metrics}

 \begin{sidewaystable}   
\centering
\caption{Selected graph metrics, $\Vert\cdot\Vert_1$ is the $L_1$ norm, $\Vert\cdot\Vert_{\ddagger}$ denotes the nuclear norm of a matrix, $\mathcal{T}(x_i)=1/|V(x_i)|\prod^{|V(x_i)|-1}_{l=1} \omega_l$}
    \label{tab:graphmetrics}

    \bigskip
    
    \begin{tabular}{lrll}
    \hline
    Distance 
    & notation & metric & level  \\
    \hline
    (normalized) Hamming 
    & $d_H(G(x_i),G(x_j))$ &  $\frac{1}{|V(x_i)|(|V(x_i)|-1)}\Vert \AM(x_i)-\AM(x_j)\Vert_1$ & local\\
     Jaccard 
     & $d_J(G(x_i),G(x_j))$ & $\frac{\Vert\AM(x_i)-\AM(x_j)\Vert_1}{\Vert\AM(x_i)-\AM(x_j)\Vert{_\ddagger}}$ & local\\
     Frobenius norm 
     & $d_F(G(x_i),G(x_j))$ &  $\sqrt{\trace((\AM(x_i)-\AM(x_j))^{\top}(\AM(x_i)-\AM(x_j)))}$& local \\
     Matrix 
     & $d_\MM(G(x_i),G(x_j))$ & $\mathrm{abs}(\MM(x_i) - \MM(x_j))$ & meso
     \\
     Matusita 
     & $d_\mathbf{W}(G(x_i),G(x_j))$ & $\sqrt{\left(\sum_{i,j}\sqrt{\mathbf{W}(x_i)}-\sqrt{\mathbf{W}(x_j)}\right)^2}$ & meso
     \\ 
    Adjacency spectral 
    & $d_\AM(G(x_i),G(x_j))$ & $\sqrt{\left(\sum_{l=1}^n\omega^{\AM(x_i)}_l-\omega^{\AM(x_j)}_l\right)^2}$ & global\\
    Laplacian spectral 
    & $d_\LM(G(x_i),G(x_j))$ & $\sqrt{\left(\sum_{l=1}^n\omega^{\LM(x_i)}_l-\omega^{\LM(x_j)}_l\right)^2}$ 
     & global\\
    normalized laplacian spectral 
    & $d_\LM^{\mathrm{norm}}G(x_i),G(x_j))$ &  $\sqrt{\left(\sum_{l=1}^n\omega^{\LM^{\mathrm{norm}}(x_i)}_l-\omega^{\LM^{\mathrm{norm}}(x_j)}_l\right)^2}$ & global \\
      Eigenspectrum distribution 
      & $d_{\varrho}(G(x_i),G(x_j))$ & $\int |\varrho(x_i)(a) -\varrho(x_j)(a)| \de a$ & global\\
        spanning tree dissimilarity  & $d_{ST}(G(x_i),G(x_j))$ &  $|\log(\mathcal{T}(x_i))-\log(\mathcal{T}(x_j))|$  & global\\
       Ipsen-Mikhailov 
       & $d_{IM}(G(x_i),G(x_j))$ & $\sqrt{\left(\int_0^{\infty}\varrho(\theta,\xi)(G(x_i))-\varrho(\theta,\xi)(G(x_j))\right)^2\de \theta} $ &  global\\
     Hamming-Ipsen-Mikhailov 
     & $d^{\xi}_{HIM}(G(x_i),G(x_j))$ & 
     $\frac{1}{\sqrt{1+\xi}}\sqrt{\mathbf{I}\MM^2+H^2}$& global \\
  \hline
    \end{tabular}
\end{sidewaystable}

\newpage

\begin{figure}
    \centering
    \includegraphics[width=0.95\textwidth]
    {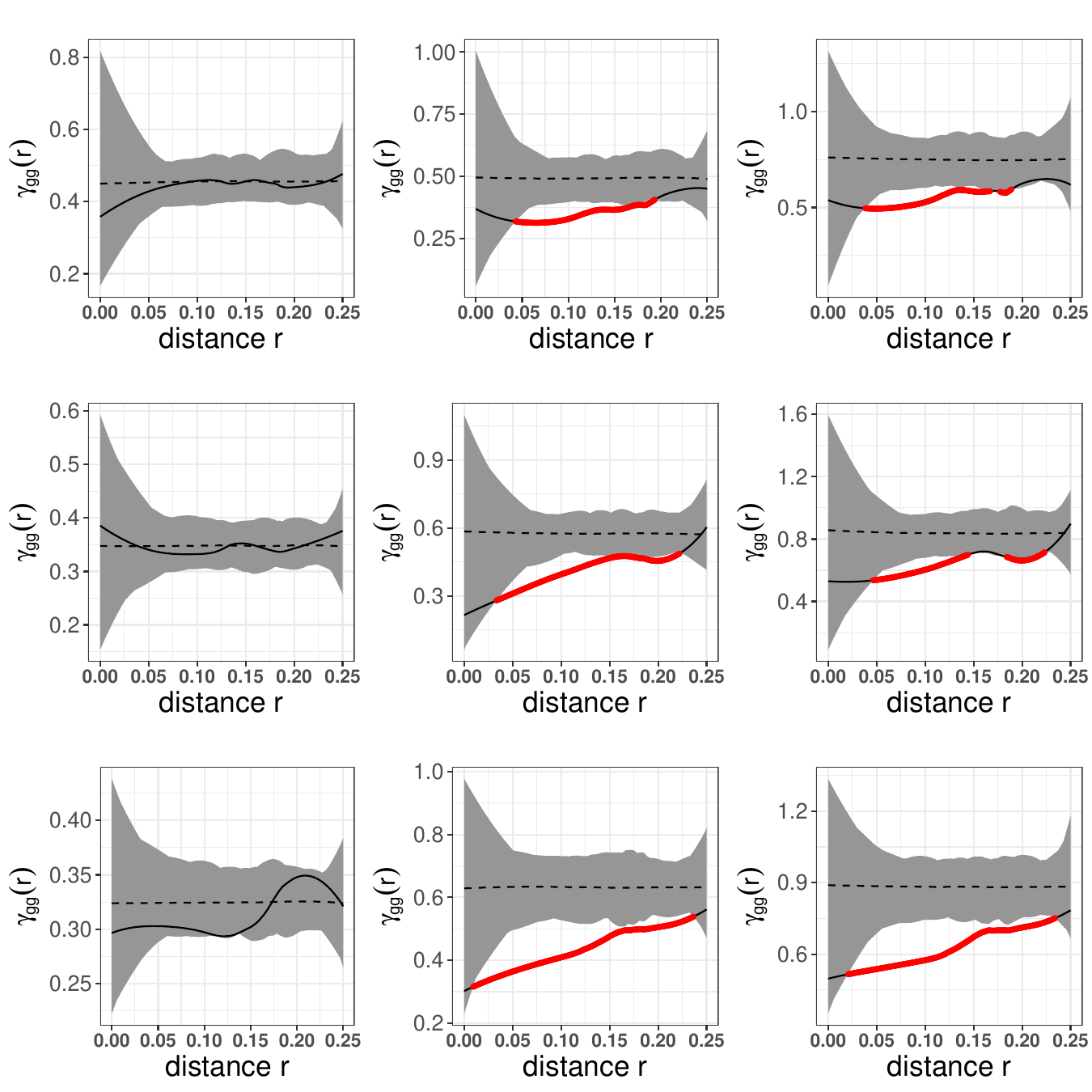}
    \caption{Graph mark variograms for simulated data from a homogenous Poisson process (top row), Strauss process (central row), and modified Thomas process (bottom row) with Erd{\"o}s-R{\'e}nyi graphs $G(n_V,p)$ with $n_V = 5$ as marks and 95\% global envelopes based on 500 random permutations of marks. From left to right: (i) graph mark variogram for $G(n_V,p)$ with  $p=0.5$, (ii) graph mark variogram for $G(n_V,p)$ with $p$ determined by the distance of the points to the boundary of the observation window with binary specified adjacency matrices, and (iii) graph mark variogram for $G(n_V,p)$ with $p$ determined by the distance of the points to the boundary of the observation window and nonzero entities of the adjacency matrices  determined by the node degrees.}
    \label{fig:SimResNv5}
\end{figure}

\label{lastpage}

\end{document}